\documentclass[3p,twocolumn,number,sort&compress]{elsarticle}

\usepackage{hyperref}
\usepackage{graphicx}
\usepackage{color}
\usepackage{colortbl}
\usepackage{rotating}
\usepackage{multirow}

\definecolor{orange}{rgb}{1,0.5,0}
\definecolor{Green}{RGB}{51,153,51}
\definecolor{Violet}{RGB}{153,0,204}
\definecolor{Blue}{rgb}{0,0,1}
\definecolor{Red}{rgb}{1,0,0}
\definecolor{Yellow}{rgb}{1,1,0}

\newcommand{\tb}{\hspace{4mm}}

\begin{document}

\title{Exploring Minimal Scenarios to Produce Transversely Bright Electron Beams Using the Eigen-Emittance Concept}
\date{\today}

\author[lanl]{L.D. Duffy}
\ead{ldd@lanl.gov}
\author[lanl]{K.A. Bishofberger}
\author[lanl]{B.E. Carlsten}
\author[um]{A. Dragt}
\author[lanl]{Q.R. Marksteiner}
\author[lanl]{S.J. Russell}
\author[lbl]{R.D. Ryne}
\author[lanl]{N.A. Yampolsky}
\address[lanl]{Los Alamos National Laboratory, Los Alamos, NM 87545, USA}
\address[lbl]{Lawrence Berkeley National Laboratory, Berkeley, CA 94720, USA}
\address[um]{Department of Physics, University of Maryland, College Park, MD 20742, USA}

\begin{abstract}

Next generation hard X-ray free electron lasers require electron beams with low transverse emittance.  One proposal to achieve these low emittances is to exploit the eigen-emittance values of the beam.  The eigen-emittances are invariant under linear beam transport and equivalent to the emittances in an uncorrelated beam.  If a correlated beam with two small eigen-emittances can be produced, removal of the correlations via appropriate optics will lead to two small emittance values, provided non-linear effects are not too large.  We study how such a beam may be produced using minimal linear correlations.  We find it is theoretically possible to produce such a beam, however it may be more difficult to realize in practice.  We identify linear correlations that may lead to physically realizable emittance schemes and discuss promising future avenues.

\end{abstract}

\begin{keyword}
emittance \sep free electron laser \sep x-rays \sep emittance exchange \sep eigen-emittance
\end{keyword}

\maketitle

\section{Introduction}

High brightness electron beams are necessary for next generation applications, including X-ray free electron lasers (XFELs), such as that of Los Alamos' Matter-Radiation Interactions in Extremes (MaRIE) experimental facility \cite{marie}.  For the specific case of the 0.25~\AA \ MaRIE XFEL, considering the beam energy spread due to single-particle synchrotron radiation limits the electron beam energy \cite{Rossbach1996401} to around 20 GeV.  Good overlap between the electron and X-ray phase spaces places a requirement on the normalized transverse emittance,  $\epsilon_n\leq\beta\gamma\lambda_x/4\pi$, where $\beta$ and $\gamma$ are, respectively, the beam's velocity normalized to the speed of light and the beam's energy normalized to the electron's rest energy, and $\lambda_x$ is the X-ray wavelength.  Thus, the beam energy limitation translates to a restriction on the transverse emittance for a given wavelength.  The normalized transverse emittances for MaRIE's XFEL must be 0.15~$\mu$m or less, while the longitudinal emittance may be much larger \cite{CarlstenJMO}.

One approach to achieve low emittance in a specific dimension is emittance partitioning.  Essentially, if the emittance in the dimension of interest is large, it may be partially transferred to a different dimension by changing the beam's initial conditions and using appropriate optics \cite{PhysRevSTAB.9.031001}.  This has been demonstrated  in the flat-beam transform (FBT) \cite{PhysRevSTAB.4.053501,PhysRevSTAB.6.104002, Carlsten:2006mn}, which partitions emittance between transverse dimensions.  A distribution of electrons with transverse correlations is produced by immersing the photo-cathode in a magnetic field.  This change in initial conditions, compared to the case with no magnetic field,  changes the beam eigen-emittances.  After these correlations are removed by appropriate beam transport, and assuming non-linear effects are small, the final emittances are equal to the beam's eigen-emittances and different to those that would occur in the absence of the applied field.  The general case of emittance partitioning between any two dimensions is discussed by Carlsten et.\ al.\ \cite{CarlstenPRSTAB}.  In addition, emittance may be exchanged between dimensions, as demonstrated in tranverse-to-longitudinal emittance exchange \cite{PhysRevSTAB.5.084001,PhysRevSTAB.14.022801,Sun:2010jj}.

Motivated by these achievements, we investigate making two transverse emittances small at the expense of the longitudinal emittance, to satisfy the emittance requirements for next generation XFELs.  Our goal is to achieve two very small emittance values by specifically tailoring the eigen-emittances \cite{PhysRevA.45.2572}.  For the purpose of this paper, \textit{emittance}  
is used to mean the root-mean-square emittance, e.g. $(mc)^{-1}\sqrt{\langle x^2\rangle\langle p_x^2\rangle - \langle x p_x\rangle^2}$ in  
the $x$-dimension, where $p_x$ is the momentum in the $x$-direction, and similarly for the other dimensions; in all cases  
when we intend \textit{eigen-emittance}, we will explicitly state `eigen-emittance'.  Eigen-emittances are conserved through linear beam transport, once the beam is initially generated.  If the beam is uncorrelated, they also correspond to the three beam emittances.  Thus, if correlations are introduced by changing the beam's initial conditions such that two of the eigen-emittances are very small, it should be possible to remove these correlations downstream and recover the eigen-emittances as the beam emittances, provided nonlinear effects are not too large.

In this work, we examine eigen-emittances produced at the cathode when the minimum number of linear correlations required to produce two small eigen-emittances are present in the electron bunch.  By first studying the eigen-emittances we aim to then tailor the beam's emittance values using this information.  Previous work has studied one specific case of correlations and examined the resulting eigen-emittances \cite{Yampolsky:2010pt}.  We use an alternative approach, in which we search for combinations of correlations that lead to our desired eigen-emittance splitting.  We aim to find the theoretically possible combinations so that we are forearmed with knowledge of which correlations are worth pursuing to obtain a transversely bright electron beam.  Future work can then focus on finding ways to realize the required correlations in physical systems.  

The correlations we consider are specifically between two of the six phase-space coordinates and mix two of the three physical dimensions, using either the coordinates themselves and/or the conjugate momenta.  Introducing a single correlation only partitions two eigen-emittance values.  As demonstrated by Yampolsky et.\ al.\ \cite{Yampolsky:2010pt} the maximum-to-minimum ratio of these two eigen-emittances always increases.  We wish to drive two eigen-emittances to small values, and require more than one correlation so that all three coordinate planes mix.  We thus require a minimum of two correlations for two small eigen-emittances, which give the minimal scenarios we examine in this work.  In section \ref{sec:theory}, we discuss the approach we have used to study the effect of correlations and the resulting eigen-emittance values.  As we are studying the least complicated scenarios, we expect these will require the simplest optics to remove the correlations and recover the beam's eigen-emittances as the rms emittance values.  In section \ref{sec:results}, we numerically investigate which combinations of correlations result in two small eigen-emittance values and classify the resulting eigen-emittance patterns for all minimal scenarios.  We find a clear prescription for the minimal correlations resulting in two small eigen-emittances, although not every combination we find has a presently recognized implementation.  We discuss the possibility of realizing the necessary correlations to obtain transversely bright electron beams in section \ref{sec:reality}.  A short conclusion is provided in section \ref{sec:discuss}.  This study provides useful guidance to exploit eigen-emittances to obtain small emittance values.  In the future, detailed analysis and numerical modelling should be performed for the schemes that show promise, however this is beyond the scope of this work.

\section{\label{sec:theory}Background Theory for Eigen-Emittances and Introducing Beam Correlations}

In this section, we briefly discuss how the beam eigen-emittances can be obtained from the beam correlation matrix and how correlations can, in theory, be introduced.  Hamiltonian motion of a beam has three moment invariants, which can conveniently be chosen to be the quantities known as the eigen-emittances.  In an uncorrelated beam, the the rms beam emittances coincide with the eigen-emittances.  A more detailed discussion of eigen-emittances is provided by Dragt, Neri and Rangarajan~\cite{PhysRevA.45.2572} and by Dragt \cite{Dragt}.   

We must work with the full six-dimensional phase-space, as we are interested in making the transverse emittances small at the expense of the longitudinal emittance.  We use canonical coordinates, $\mathbf{s}=(x, p_x, y, p_y, z, p_z)^T$, where $p_x$, $p_y$ and $p_z$ are the canonically conjugate momenta to the configuration space coordinates, $x$, $y$ and $z$, respectively.  The superscript, $T$, denotes transpose, i.e.\ $\mathbf{s}$ is defined as a \textit{column} vector.  For simplicity, we choose the coordinates to be dimensionless, defined in the same manner as used in Carlsten et.\ al.\ \cite{CarlstenPRSTAB}, i.e.\ as the dimensionless deviations from a reference trajectory, $\mathbf{s}_t$,
\begin{eqnarray}
(x-x_t)/l \mapsto x , &\qquad & (p_x-p_{xt})/k \mapsto p_x \nonumber \\ 
(y-y_t)/l \mapsto y , &\qquad & (p_y-p_{yt})/k \mapsto p_y \\\ 
(z-z_t)/l \mapsto z  ,&\qquad & (p_z-p_{zt})/k \mapsto p_z , \nonumber 
\end{eqnarray}
where $l$ and $k$ are scaling factors for the position and momentum coordinates, respectively.  We leave these scale factors undefined, as we are simply interested in how the eigen-emittance values change as correlations are introduced and increased.  The conversion from canonical coordinates to more typically used coordinates, such as time and energy, are discussed in detail by Carlsten et.\ al.\ \cite{CarlstenPRSTAB}.  In this work, we choose canonical coordinates out of convenience.

From the particle phase-space coordinates, we can construct the beam matrix, $\Sigma=\sum_{i=1}^N\mathbf{s}_i\mathbf{s}_i^T/N$, where $N$ is the total number of particles and the sum is over the particles.   The definition of the eigen-emittances arises from the fact that $\Sigma$ can be transformed, via a symplectic congruency transformation, $R$, to the form, $\Sigma_{e}\equiv R\Sigma R^T = \textrm{diag}(\lambda_1,\lambda_1,\lambda_2,\lambda_2,\lambda_3,\lambda_3)$.  The quantities $\lambda_j$ are the eigen-emittances.  In practice, these quantities can be obtained from the beam matrix as the absolute value of the eigenvalues of the characteristic equation,  $det(J\Sigma-i\lambda_j I)=0$, where $I$ is the identity matrix and the only non-zero entries in the matrix, $J$, are the $2\times2$ block diagonal entries that contain the skew-symmetric matrix, 
\begin{equation}
J_2=\left(\begin{array}{cc}
0 & 1 \\
-1 & 0
\end{array}\right) \; .
\end{equation}
$I$ and $J$ have the same dimensionality as the phase-space.  The eigen-emittances are invariant under a symplectic transformation, i.e.\ under a matrix transformation, $M$, of $\Sigma$, $M\Sigma M^T$ where $M$ satisfies the symplectic condition, $MJM^T=J$.  Only non-symplectic linear transformations can alter the eigen-emittances.

To introduce correlations in the initial conditions, we transform an initially uncorrelated beam with beam matrix, $\Sigma_0=\mathrm{diag}(\sigma_{x}^2, \sigma_{p_x}^2, \sigma_y^2, \sigma_{p_y}^2, \sigma_z^2, \sigma_{p_z}^2)$, via
\begin{eqnarray}
\Sigma&=&\frac1N\sum_{i=1}^N\mathbf{s_i}\mathbf{s}_i^T\nonumber \\
&=&\frac1N\sum_{i=1}^N(I+C)\mathbf{s}_{0i}\mathbf{s}_{0i}^T(I+C)^T \nonumber\\
&\equiv&(I+C)\Sigma_0(I+C)^T \; ,
\label{eq:sigtrans}
\end{eqnarray}
where the matrix, C, which we shall refer to as the \textit{C-matrix}, contains the part of the transformation which introduces correlations, i.e.\ the final variables, $\mathbf{s}$, after correlations are introduced are a function of the initial variables, $\mathbf{s}_0$, via $\mathbf{s}=(I+C)\mathbf{s_0}$.  The transformations we study preserve the beam brightness, or equivalently, the six-dimensional phase space volume, i.e.\ we ensure $\mathrm{det}(I+C)=1$.  $C$ describes the change in initial conditions.  The C-matrix is not symplectic, i.e.\ $CJC^T\neq J$ and thus the transformation of Eq.~(\ref{eq:sigtrans}) changes the eigen-emittances.  Given that we are interested in correlations between different physical dimensions only , the most general form of our C-matrix is 
\begin{equation}
C=\left( \begin{array}{cccccc}
0& 0&c_{13}&c_{14}&c_{15}&c_{16} \\
0 & 0&c_{23}&c_{24}&c_{25}&c_{26} \\
c_{31}&c_{32}& 0 & 0 & c_{35}&c_{36} \\
c_{41}&c_{42}& 0 & 0 & c_{45}&c_{46} \\
c_{51}&c_{52}& c_{35}&c_{36} & 0 & 0 \\
c_{61}&c_{62}& c_{35}&c_{36}& 0 & 0 
\end{array}\right) \, ,
\label{eq:C}
\end{equation}
i.e. the matrix has $2\times2$ zero blocks on the diagonal and the correlations mix phase-space components between the $x$, $y$ and $z$ dimensions.  The zero blocks occur due to the fact that we are not interested in correlations between the coordinates and their conjugate momenta for this study.  As we are examining minimal cases, the majority of the entries in the C-matrices we use will be zero, however all $c_{jk}$ in (\ref{eq:C}) have the potential to be non-zero.  If we wish to introduce a single correlation, then only one of the $c_{jk}$ is non-zero.  

If we allow completely general C-matrices, any eigen-emittance combination that we desire, such as two or even three very small values, can be at least theoretically obtained.  This follows from the fact that for any positive definite symmetric matrix, $Z$, by the Law of Inertia for quadratic forms \cite{QuadraticForm}, there is an invertible matrix, $G$, such that $GZG^T=\mathrm{diag}(\mu_1, \mu_2,...,\mu_6)$ and the $\mu_j$ are \textit{any} prescribed positive numbers.  However, we are restricting our study to a specific form of the C-matrix with few non-zero entries, corresponding to the introduction of two linear correlations at the cathode.  Our concern is to identify the simplest cases that produce our desired result and are worth further investigation.

Two correlations may be introduced to a beam via
\begin{eqnarray}
\Sigma & = &(I+C_2)(I+C_1)\Sigma_0(I+C_1)^T(I+C_2)^T \label{eq:depcorr} \\
&\equiv&(I+C)\Sigma_0(I+C)^T\nonumber\; ,
\end{eqnarray}
where $C_1$ and $C_2$ each represent a single correlation and thus contain one non-zero entry each.  An important point in introducing two correlations to a beam is whether the correlations are independent of each other or not.  
If $C_1$ and $C_2$ commute, then the introduced correlations are independent.  However, if $C_1$ and $C_2$ do not commute then $(I+C_1)(I+C_2)\neq(I+C_2)(I+C_1)$, i.e. the order in which the correlations are introduced becomes important.  In the case that they are independent, the resulting C-matrix will have only two non-zero entries.  If they are dependent, the C matrix will contain three non-zero entries.  Finally, we note that products of matrices of the form, $(I+C_j)$, begin to approximate arbitrary matrices, $G$, and therefore, in principle, can lead to lowering all three eigen-emittances.  

When we discuss specific correlations in this paper, the final variable is the first in the pair and the initial variable that introduces the correlation is the second, e.g. if only an $x$-$y$ correlation is used, the functional dependence is $x(x_0,y_0)$.  An example of two correlations that do not commute is $x$-$y$ and $y$-$z$.  If $x$-$y$ is introduced to the beam first, then the functional dependence is $x(x_0,y_0)$ and $y(y_0,z_0)$ and the correlations are independent.  If, however, the $y$-$z$ correlation is introduced first, the functional dependence is $y(y_0,z_0)$ and $x(x_0,y)\equiv x(x_0,y(y_0,z_0))$ and the correlations are dependent.  It is easily demonstrated for this case that the resulting C-matrix depends on the order in which these correlations are applied.  To make functional dependence of variables clear from  the dependence or independence of correlations, throughout this paper we refer to the variables as either initial or final variables.  

In the following section, the effects of introducing different pairs of correlations on the eigen-emittances are investigated, using numerical variation of entries in the C-matrix with \textit{Mathematica} \cite{Mathematica7}.  
  
\section{\label{sec:results}Change in Eigen-emittances Due to Initial Correlations}

Our goal is to find pairs of correlations which decrease two eigen-emittances at the expense of the third.  To see the effect of the correlations, we enter the appropriate correlations in the C-matrix, solve for the eigen-emittances and see how these results vary with increasing correlation strength.  In the case of dependent correlations, the final C-matrix is calculated using Eq.~(\ref{eq:depcorr}).  For numerical purposes, we begin with initial beam emittances 0.7/0.7/1.4 for the $x$/$y$/$z$ emittances, which correspond to the eigen-emittances in the uncorrelated beam.  If the units are chosen so that the emittances are in microns, these are typical emittances for a 500~pC photo-injector.   

\begin{figure}
\begin{center}
\includegraphics[clip=false,keepaspectratio=true,trim=0cm 0 0cm 0,height=0.22\textheight]{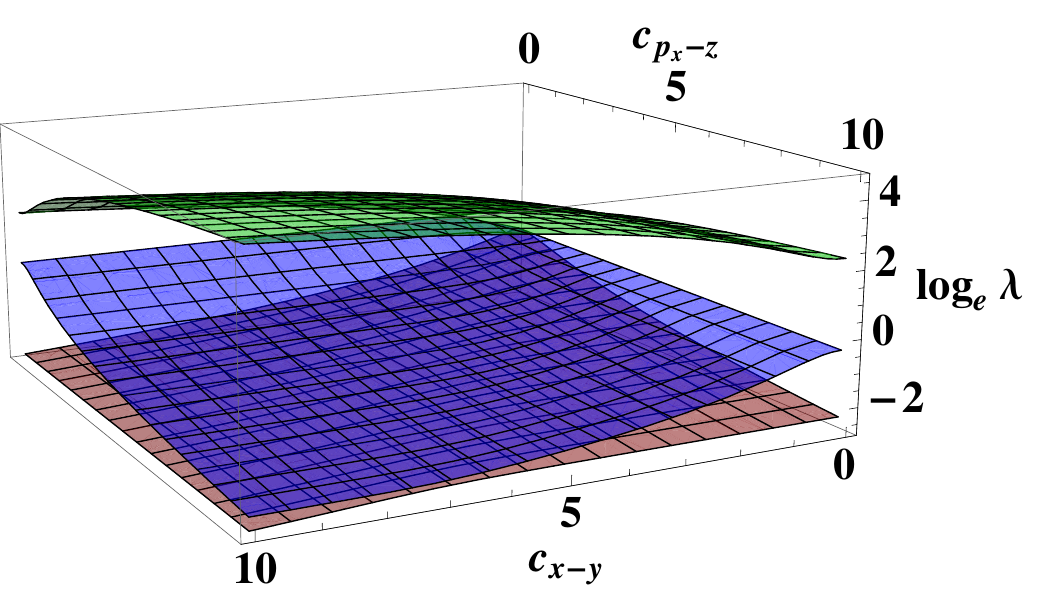}\\
Case (1)\\
\vspace{5mm}
\includegraphics[clip=false,keepaspectratio=true,trim=3cm 0cm 3cm 0cm,height=0.22\textheight]{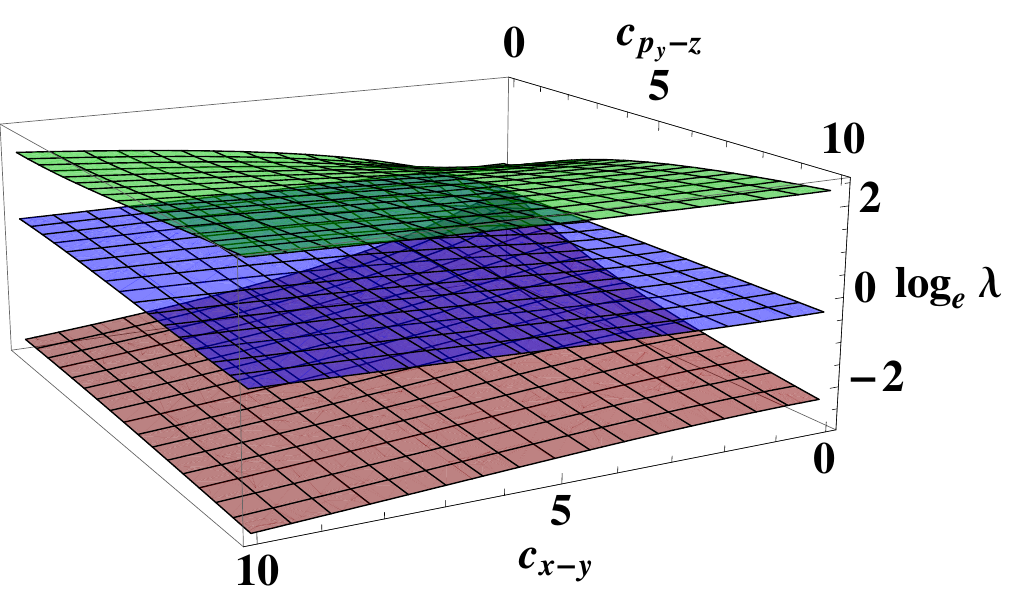}\\
Case (2)\\
\vspace{5mm}
\includegraphics[clip=false,keepaspectratio=true,trim=0cm 0cm 0cm 0cm,height=0.22\textheight]{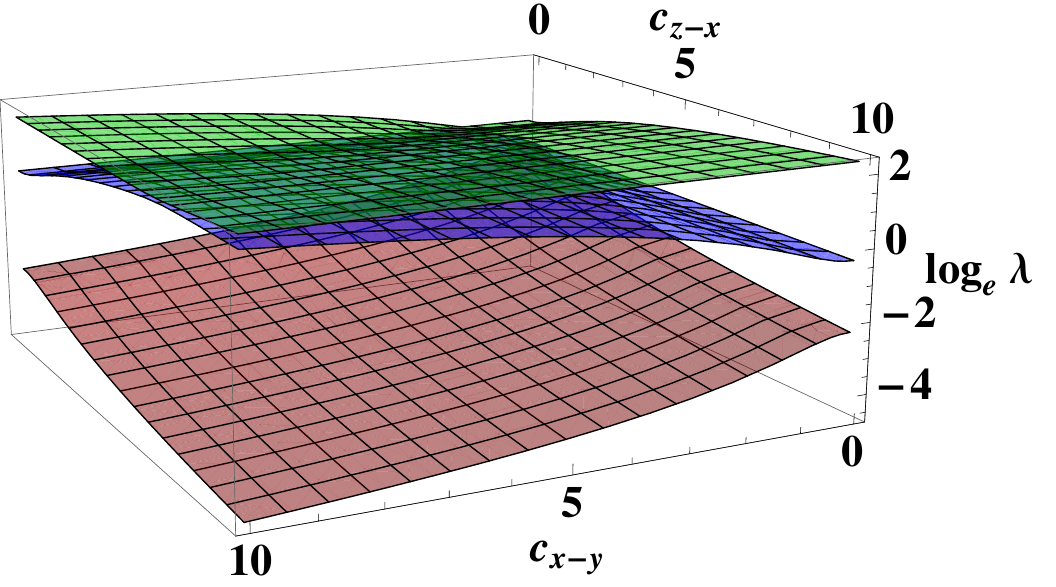}\\
Case (3)\\
\end{center}
\caption{\label{fig:eemitsplit}  Three representative cases of the variation of the eigen-emittance values, $\lambda$, with two correlations.  Case (1):  (top) $x$-$y$ and $p_x$-$z$ correlations give two small and one large eigen-emittance.  Case (2):   (center) $x$-$y$ and $p_y$-$z$ correlations give one large, one small and one relatively unchanged eigen-emittance.  Case (3):  (bottom) Independent $x$-$y$ and $z$-$x$ correlations result in two large and one small eigen-emittance.}
\end{figure}

We find that the results fall under three cases, as shown in Fig.~\ref{fig:eemitsplit}: (1) our desired outcome of two small and one large eigen-emittance; (2) one small eigen-emittance value, one large, and one relatively unchanged; or (3)  two large eigen-emittance values and one small.  These results are not completely unexpected.  As mentioned previously, Yampolsky et.\ al.\ proved that for a single correlation in a two-dimensional system, corresponding to one entry in the C-matrix, the maximum-to-minimum eigen-emittance ratio always increases \cite{Yampolsky:2010pt}.  It is reasonable to expect that this structure is still present when two correlations are introduced, and it is also seen in Fig.~\ref{fig:eemitsplit} along each axis (i.e.\ where only one correlation remains).  The remaining question is then what happens to the third eigen-emittance value.  Does it increase, decrease or remain relatively unchanged?  As shown in Fig.~\ref{fig:eemitsplit}, all of these scenarios are possible.

We have completely classified the combinations of two linear correlations leading to each case, using numerical evaluation.  The number of cases to be evaluated can be reduced by recognizing symmetries in the problem, such as the cyclic permutations of the coordinates.  It is easiest to classify the cases if we first consider which combinations of correlations are possible.  As we are required to introduce correlations between all three dimensions, variables from one of the dimensions must appear twice, once in each pair of correlations.  These can occur as either a repeated phase-space coordinate (either position or momentum) or a canonically conjugate pair.  We first consider the independent correlations.  For these cases, the two variables both associated with one dimension can occur in one of three ways:  both are the initial variable, both are the final variable or one is an initial variable and one is a final variable.  The eigen-emittance split obtained for each of these six cases is shown in Table~\ref{tab:eclass}.

\begin{table}
\begin{center}
\begin{tabular}{lc} \hline
Variables&Resulting case\\
 \hline\hline
 (1) 2 repeated &  \\
 \tb (a) both initial & 2 \\
 \tb (b) 1 initial, 1 final & 3 \\
 \tb (c) both final & 2 \\
 (2) 2 conjugate &  \\
 \tb (a) both initial & 3 \\
 \tb (b) 1 initial, 1 final & 2 \\
 \tb (c) both final & 1 \\ \hline
\end{tabular}
\end{center}
\caption{\label{tab:eclass}Independent correlations leading to the particular cases in Fig.~\ref{fig:eemitsplit}, which are: (1) two small and one large eigen-emittance, (2) one small, one large and one relatively unchanged eigen-emittance, and (3) two large and one small eigen-emittance.  The two variables refer to the variables of the two correlations that are associated with the same dimension.}
\end{table}

As shown in Table~\ref{tab:eclass}, independent correlations in which the final variables are canonically conjugate lead to two small eigen-emittance values.  The exact combinations of independent correlations that result in two small eigen-emittances are illustrated in Fig.~\ref{fig:colorchart}.  Additionally, all \textit{dependent} correlations lead to two small eigen-emittance values.  These cases leading to two small eigen-emittances have recently been presented \cite{PAC11}, however this is the first time the full classification of the minimal linear correlation scenarios has been discussed.

\begin{figure}
\begin{center}

\begin{tabular}{|c|c|c|c|c|c|c|c|c|}\cline{4-9}
\multicolumn{1}{c}{} &\multicolumn{1}{c}{} & &\multicolumn{6}{|c|}{Column Index}\\
\cline{4-9}
\multicolumn{1}{c}{}&\multicolumn{1}{c}{} & &$x_0$& $p_{x0}$ & $y_0$ & $p_{y0}$ & $z_0$ & $p_{z0}$\\
\cline{4-9} 
\multicolumn{1}{c}{}&\multicolumn{1}{c}{} & &$1$& $2$ &$3$ &$4$ &$5$ &$6$\\
\hline
\multirow{6}{*}{\begin{sideways}Row Index\end{sideways}}
&$x$& $1$&\multicolumn{2}{|c|}{\cellcolor[gray]{0.0}} &\multicolumn{2}{|c|}{\cellcolor{Yellow}A} 
	&\multicolumn{2}{|c|}{\cellcolor{Red}\color{white}B} 
	\\ 
\cline{2-9} 
&$p_x$ 	&$2$ & \multicolumn{2}{|c|}{\cellcolor[gray]{0.0}}&\multicolumn{2}{|c|}{\cellcolor{Red}\color{white}B} &\multicolumn{2}{|c|}{\cellcolor{Yellow}A}\\ \cline{2-9} 
&$y$ 	&$3$&\multicolumn{2}{|c|}{\cellcolor{orange}C}&\multicolumn{2}{|c|}{\cellcolor[gray]{0.0}}&\multicolumn{2}{|c|}{\cellcolor{Blue}\color{white}D} \\ \cline{2-9}
&$p_y$	&$4$ &\multicolumn{2}{|c|}{\cellcolor{Blue}\color{white}D} &\multicolumn{2}{|c|}{\cellcolor[gray]{0.0}} &\multicolumn{2}{|c|}{\cellcolor{orange}C} \\ \cline{2-9} 
&$z$	&$5$ & \multicolumn{2}{|c|}{\cellcolor{Violet}\color{white}E} &\multicolumn{2}{|c|}{\cellcolor{Green}\color{white}F}  &\multicolumn{2}{|c|}{\cellcolor[gray]{0.0}} \\ \cline{2-9}
&$p_z$&$6$&\multicolumn{2}{|c|}{\cellcolor{Green}\color{white}F}  &\multicolumn{2}{|c|}{\cellcolor{Violet}\color{white}E}  &\multicolumn{2}{|c|}{\cellcolor[gray]{0.0}} \\
\hline
\end{tabular}

\end{center}
\caption{\label{fig:colorchart}Chart of independent correlations leading to two small eigen-emittances.  Two entries in the C-matrix must be chosen, one from each matching block, i.e. blocks labelled by the same letter.  The blank blocks are not considered, as they do not correlate two of the three dimensions.}
\end{figure}

We have now classified all combinations of two correlations and the pattern of their resulting eigen-emittance spectrum.  In the following section, we discuss the possibility of physically realizing a combination leading to two small eigen-emittance values at the cathode.


\section{\label{sec:reality}Physical Realization of Correlations}

While we have found that combinations of two correlations can lead to two small eigen-emittance values, it is not as simple to find a scheme that may be physically realized.  The correlations that are currently known to be able to be introduced at the cathode are:
\begin{itemize}
\item $x$-$y$ correlations using an elliptical cathode or laser spot;
\item $z$-$x$ or $z$-$y$ correlations using a drive laser with a tilted pulse front, as in the scheme described in Yampolsky et.\ al.\ \cite{Yampolsky:2010pt} or using a tilted cathode;
\item $x$-$z$ or $y$-$z$ by scanning a drive laser across the cathode;
\item $p_z$-$x$ or $p_z$-$y$ by scanning a drive laser with a frequency modulation across the cathode or using a photo-cathode with changing work function as suggested in Carlsten et.\ al.\ \cite{CarlstenPRSTAB}; and
\item $p_x$-$y$ and $p_y$-$x$ together as angular momentum by applying a magnetic field perpendicular to the cathode.
\end{itemize}

The $p_x$-$y$ and $p_y$-$x$ are simple to introduce together at the cathode using a solenoid, however they do not lead to two small eigen-emittances by themselves.  Initial investigation of combinations containing both these correlations and a third indicate that they do not lead to two small eigen-emittances, and support the findings of Yampolsky et.\ al.\ \cite{Yampolsky:2010pt} which studied these correlations combined with what is effectively a $z$-$x$ correlation.

From the list above, the methods $p_z$-$x$ or $p_z$-$y$ correlations will also introduce correlations between the coordinate values.  However, these correlations do not couple all three dimensions and will not lead to two small eigen-emittance values.  It may be possible to still obtain two small eigen-emittance using $p_z$-$x$ or $p_z$-$y$ with $z$-$y$ or $z$-$x$, respectively, despite the presence of additional correlations.  An initial study of more than two correlations shows that it is at least possible in some cases.  

Theoretically and for completeness, we have included correlations depending on the beam momenta, but no practical implementation currently exists.   Also, a $p_y$-$z$ or $p_x$-$z$ correlation would be difficult to create at the cathode.

Given the above, the only possible independent correlations that might be physically realizable are the purple and green blocks of Fig.~\ref{fig:colorchart} and combinations involving two dependent correlations.  From the correlations we can implement, listed at the start of this discussion, the scenarios that may be possible are:
\begin{itemize}
\item independent $p_z$-$x$/$y$ with $z$-$y$/$x$ correlations;
\item dependent correlations between all three position coordinates; and
\item dependent $p_z$-$x$/$y$ with the appropriate $x$-$y$ or $y$-$x$ correlations.
\end{itemize}

For purely linear transport, these correlations must be introduced when the electron bunch is formed at the cathode.  Once the beam is created, the eigen-emittances may only be changed by non-symplectic effects.  This offers both a difficulty and an opportunity.  Should we succeed in creating a beam with the correlations discussed above, non-linear effects may alter the eigen-emittance values and should be minimized in the beam line where possible.  However, there is an opportunity to use an alternative approach, introducing correlations downstream using non-symplectic transport, such as by introducing a correlation between the energy and transverse position using a tapered foil \cite{4332828,CarlstenPRSTAB}.  The phase-space volume is not expected to be preserved in such a scenario and these type of transformations are not covered by this work.

\section{\label{sec:discuss}Conclusion}

We have classified the eigen-emittance spectra for minimal linear beam correlations introduced at the cathode, identifying the cases which promise to lead to two small eigen-emittance values.  Further investigation into these cases is warranted, to make sure they are able to be implemented practically and that any additional correlations that may be introduced while creating these beams does not change the eigen-emittance values by too much.  We have also found the minimal cases that do not give two small eigen-emittance values, providing a guide to future studies seeking to exploit eigen-emittance values to produce bright electron beams.

Should the cases in this work not lead to an implementable scheme that, after removal of correlations, gives a transversely bright electron beam, more sophisticated combinations of correlations may still be viable.  However, it is expected that more complicated optics would be required to transform such a beam to the uncorrelated case.

In this study, we have not included the effects of correlations between the position and momentum coordinates in a single dimension, as we were searching for correlations between dimensions that result in two small eigen-emittances.  The effects of these correlations should be investigated to make sure that they do not alter the eigen-emittance values significantly in our cases of interest.  Additionally, the magnitude of the required correlations in physically possible schemes leading to two small eigen-emittances should be further investigated, to make sure that they are not too large to be practically realized.

\section*{Acknowledgments}
We acknowledge the support of the U.S. Department of Energy through the Laboratory Directed Research and Development program at Los Alamos National Laboratory.

\bibliographystyle{elsarticle-num}
\bibliography{eemit}

\end{document}